\documentclass[a4paper]{article}
\usepackage{epsfig,amsmath,graphicx,amssymb,xcolor,color,dsfont}
\usepackage{INTERSPEECH2020}
\usepackage[linesnumbered,ruled,vlined]{algorithm2e}
\SetArgSty{textnormal}
\usepackage[hidelinks]{hyperref}
\usepackage{url}
\usepackage{booktabs}
\usepackage{multirow}

\newcommand{\transpose}{^{^{\intercal}}}
\ninept 
\title{Neural PLDA Modeling for End-to-End  Speaker Verification}
\name{Shreyas Ramoji, Prashant Krishnan, Sriram Ganapathy
\thanks{This work was funded by the Ministry of Human Resources Development (MHRD) of India and the Department of Science and Technology (DST) EMR/2016/007934 grant.}
\address{Learning and Extraction of Acoustic Patterns (LEAP) Lab, \\Department of Electrical Engineering, Indian Institute of Science, Bengaluru, India}
\email{\{shreyasr, prashantkv1, sriramg\}@iisc.ac.in}}
\begin{document}

\maketitle

\begin{abstract}
While deep learning models have made significant advances in supervised classification problems, the application of these models for out-of-set verification tasks like speaker recognition has been limited to deriving feature embeddings. The state-of-the-art x-vector PLDA based speaker verification systems use a generative model based on probabilistic linear discriminant analysis (PLDA) for computing the verification score. Recently, we had proposed a neural network approach  for backend modeling in speaker verification called the neural PLDA (NPLDA) where the likelihood ratio score of the generative PLDA model is posed as a discriminative similarity function and the learnable parameters of the score function are optimized using a verification cost. In this paper, we extend this work to achieve joint optimization of the embedding neural network (x-vector network) with the NPLDA network in an end-to-end (E2E) fashion. This proposed end-to-end model is optimized directly from the acoustic features with a verification cost function and during testing, the model directly outputs the likelihood ratio score. With various experiments using the NIST speaker recognition evaluation (SRE) 2018 and 2019 datasets, we show that the proposed E2E model improves significantly over the  x-vector PLDA baseline speaker verification system.
\end{abstract}

\section{Introduction}
Automatic speaker verification (ASV) has several applications such as voice biometrics for commercial applications, speaker detection in surveillance, speaker diarization, etc. A speaker is enrolled by a sample utterance(s), and the task of ASV is to detect whether the target speaker is present in a given test utterance or not. Several challenges have been organized over the years for benchmarking and advancing speaker verification  technology such as the NIST speaker recognition Evaluation (SRE) challenge 2019 \cite{Sadjadi19plan}, the VoxCeleb speaker recognition challenge (VoxSRC) \cite{nagrani2017voxceleb} and the VOiCES challenge \cite{Nandwana2019}.
The major challenges in speaker verification include the language mismatch in testing, short duration audio and the presence of noise/reverberation in the speech data. 

The state-of-the-art systems in speaker verification use a model to extract embeddings of fixed dimension from utterances of variable duration. The earlier approaches based on unsupervised Gaussian mixture model (GMM) i-vector extractor \cite{dehak2010front} have been recently replaced with neural embedding extractors \cite{snyder2016deep,snyder2018x}  which are trained on large amounts of supervised speaker classification tasks.  These fixed dimensional embeddings are pre-processed with a length normalization \cite{garcia2011analysis} technique followed by probabilistic linear discriminant analysis (PLDA) based backend modeling approach \cite{kenny2010bayesian}. 

In our previous work, we had explored a discriminative neural PLDA (NPLDA) approach \cite{ramoji2020pairwise} to backend modeling where  a discriminative similarity function was used. The learnable parameters of the NPLDA model were optimized using an approximation  of  the  minimum  detection  cost  function (DCF). This model also showed good improvements in our SRE evaluations and the VOiCES from a distance challenge \cite{ramoji2020leap, ramoji2020nplda}. In this paper, we extend this work to propose a joint modeling framework that optimizes both the front-end x-vector embedding model and the backend NPLDA model in a single end-to-end (E2E) neural framework. The proposed model is initialized with the pre-trained x-vector time delay neural network (TDNN). The NPLDA E2E is fully trained on pairs of speech utterances starting directly from the mel-frequency cepstral coefficient (MFCC) features. The advantage of this method is that both the embedding extractor as well as the final score computation is optimized on pairs of utterances and with the speaker verification metric. With experiments on the NIST SRE 2018 and 2019 datasets, we show that the proposed NLPDA E2E model improves significantly over the baseline system using x-vectors and generative PLDA modeling.    



\section{Related Prior Work}
The common approaches for scoring in speaker verification systems include support vector machines (SVMs) \cite{campbell2006support}, and the probabilistic linear discriminant analysis (PLDA) \cite{kenny2010bayesian}. Some efforts on pairwise generative and discriminative modeling are discussed in \cite{cumani2013pairwise,cumani2014large,cumani2014generative}. The discriminative version of PLDA with logistic regression and support vector machine (SVM) kernels has also been explored in ~\cite{burget2011discriminatively}. In this work, the authors use the functional form of the generative model and pool all the parameters needed to be trained into a single long vector. These parameters are then discriminatively trained using the SVM loss function with pairs of input vectors. The discriminative PLDA (DPLDA) is however prone to over-fitting on the training speakers and leads to degradation on unseen speakers in SRE evaluations~\cite{villalba2020state}. The regularization of embedding extractor network using a Gaussian backend scoring  has been investigated in \cite{Ferrer2019}.
Other recent developments in this direction includes efforts in using the approximate DCF metric for text dependent speaker verification \cite{Mingote2019}.

\begin{figure*}[t]
    \centering
    \includegraphics[width=0.9\textwidth,trim={1.5cm 0.3cm 1.5cm 0.5cm},clip]{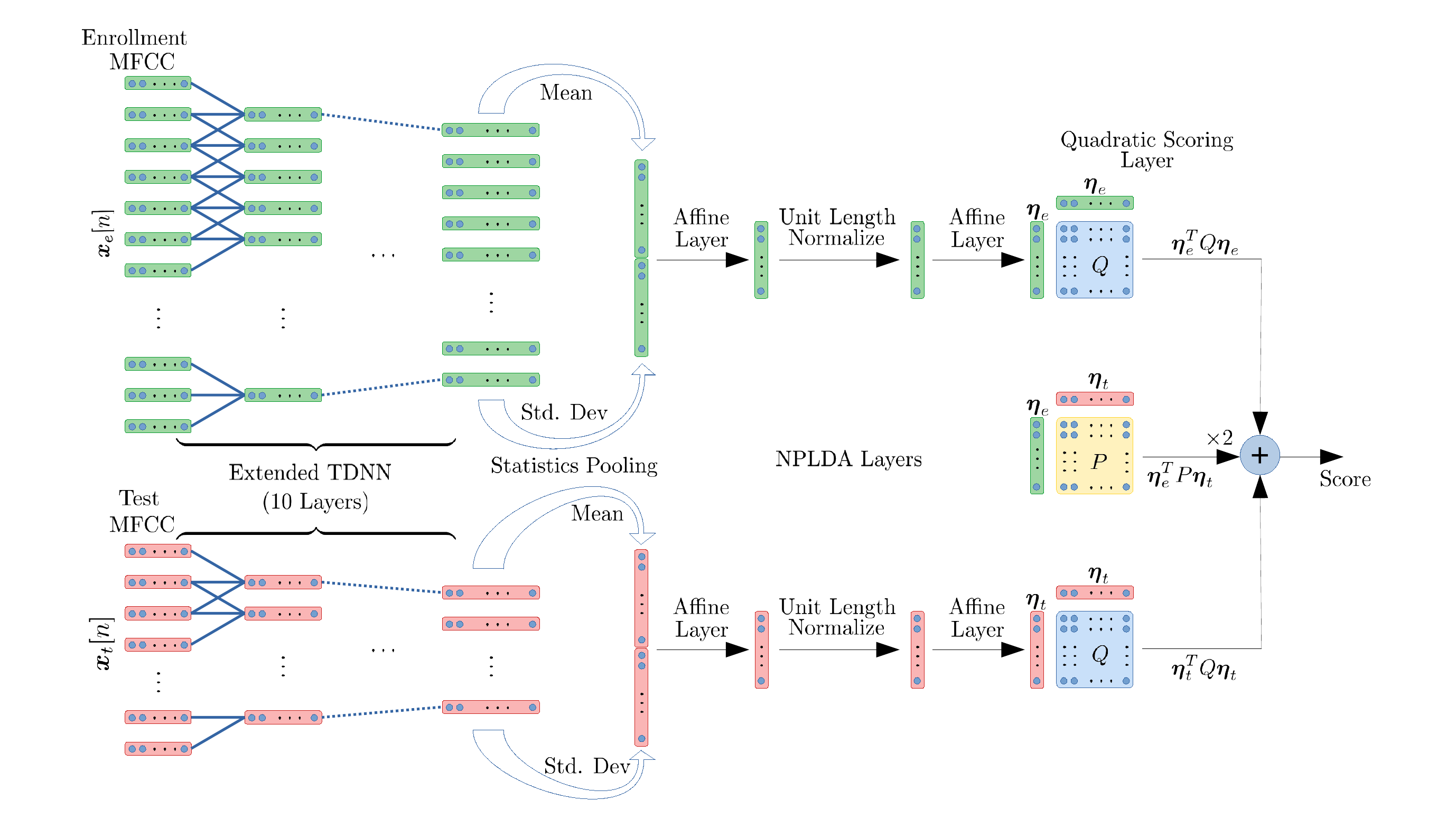}
    \caption{End-to-End x-vector NPLDA architecture for Speaker Verification.}
    \label{fig:e2e}
\end{figure*}

Recently, some end-to-end approaches for speaker verification have been examined. For example, in~\cite{rohdin2018end}, the PLDA scoring which is done with the i-vector extraction has been jointly derived using a deep neural network architecture and the entire model is trained using a binary cross entropy training criterion. In \cite{wan2018generalized}, a generalized end to end loss by minimizing the centroid means of within speaker distances while maximizing across speaker distances was proposed. In another E2E effort, the use of triplet loss has been explored \cite{Zhang2017}. However, in spite of these efforts, most state of the art systems use a generative PLDA backend model with x-vectors and similar neural network embeddings.

\section{Background}

\subsection{Generative Gaussian PLDA (GPLDA) }
The PLDA model on the processed x-vector embedding $\boldsymbol{\eta} _r$ (after centering, LDA transformation and unit length normalization) is given by
\begin{equation}
\boldsymbol{\eta} _r = \Phi \boldsymbol{\omega} + \boldsymbol{\epsilon}_r 
\end{equation}
where $\boldsymbol{\omega}$ is the latent speaker factor with a Gaussian prior of $\mathcal{N}(\textbf{0},\textbf{I})$, $\Phi$ characterizes the speaker sub-space matrix, and $\boldsymbol{\epsilon}_r$ is the residual assumed to have distribution $\mathcal{N}(\textbf{0},\boldsymbol{\Sigma})$. 
For scoring, a pair of embeddings, $\boldsymbol{\eta}_e$ from the enrollment recording and $\boldsymbol{\eta}_t$ from the test recording are used with the PLDA model to compute the log-likelihood ratio score given by
\begin{equation}\label{eq:plda_scoring}
s(\boldsymbol{\eta}_e, \boldsymbol{\eta}_t) = \boldsymbol{\eta}_e\transpose \boldsymbol{Q}\boldsymbol{\eta}_e + \boldsymbol{\eta}_t\transpose \boldsymbol{Q}\boldsymbol{\eta}_t + 2\boldsymbol{\eta}_e\transpose \boldsymbol{P}\boldsymbol{\eta}_t + \text{const}
\end{equation}
where,
\begin{eqnarray}
\boldsymbol{Q} & = & \boldsymbol{\Sigma} _{tot} ^{-1} -  (\boldsymbol{\Sigma} _{tot} - \boldsymbol{\Sigma} _{ac} \boldsymbol{\Sigma} _{tot}^{-1} \boldsymbol{\Sigma} _{ac})^{-1} \\
\boldsymbol{P} &  = &  \boldsymbol{\Sigma} _{tot} ^{-1} \boldsymbol{\Sigma} _{ac} (\boldsymbol{\Sigma} _{tot} - \boldsymbol{\Sigma} _{ac} \boldsymbol{\Sigma} _{tot}^{-1} \boldsymbol{\Sigma} _{ac})^{-1}
\end{eqnarray}
with $\boldsymbol{\Sigma} _{tot} = \Phi \Phi ^T + \boldsymbol{\Sigma}$ and $\boldsymbol{\Sigma} _{ac} = \Phi \Phi ^T$. 

In the kaldi implementation of PLDA, a diagonalizing transformation which simultaneously diagonalizes the within and between speaker covariances is computed which reduces $\boldsymbol{P}$ and $\boldsymbol{Q}$ to diagonal matrices.

\pagebreak
\subsection{NPLDA}\label{sec:PldaNet}
In the discriminative NPLDA approach \cite{ramoji2020nplda}, we construct the pre-processing steps of LDA as first affine layer, unit-length normalization as a non-linear activation and PLDA centering and diagonalization as another affine transformation. The final PLDA pair-wise scoring given in Eq.~\ref{eq:plda_scoring} is implemented as a   quadratic layer in Fig.~\ref{fig:e2e}. Thus, the NPLDA implements the pre-processing of the x-vectors and the PLDA scoring as a neural backend. 

\subsubsection{Cost Function}\label{sec:costfuncs}
To train the NPLDA for the task of speaker verification, we sample pairs of x-vectors representing target (from same speaker) and non-target hypothesis (from different speakers). The normalized detection cost function (DCF) \cite{van2007introduction} for a detection threshold $\theta$ is defined as:
\begin{align}\label{eq:det_cost}
    C_{Norm}(\beta,\theta) = P_{Miss}(\theta) + \beta P_{FA}(\theta)
\end{align}
where $\beta$ is an application based weight  defined as
\begin{align}
    \beta = \frac{C_{FA} (1-P_{target})}{C_{Miss}P_{target}}
\end{align}
where $C_{Miss}$ and $C_{FA}$ are the costs assigned to miss and false alarms, and $P_{target}$ is the prior probability of a target trial.
$P_{Miss}$ and $P_{FA}$ are the probability of miss and false alarms respectively, and are computed by applying a detection threshold of $\theta$ to the log-likelihood ratios.
A differentiable approximation of the normalized detection cost was proposed in \cite{ramoji2020nplda, Mingote2019}.  
\begin{align}
    P_{Miss}^{\text{(soft)}}(\theta) &= \frac{\sum_{i=1}^{N} t_i  \left[1-{\sigma}(\alpha(s_i-\theta))\right]}{\sum_{i=1}^{N} t_i} \\
    P_{FA}^{\text{(soft)}}(\theta) &= \frac{\sum_{i=1}^{N} (1-t_i) {\sigma}(\alpha(s_i - \theta))}{\sum_{i=1}^{N} (1-t_i)}
\end{align}

\noindent Here, $i$ is the trial index, $s_i$ is the system score and $t_i$ denotes the ground truth label for trial $i$, and $\sigma$ denotes the sigmoid function. $N$ is the total number of trials in the minibatch over which the cost is computed. By choosing a large enough value for the warping factor $\alpha$, the approximation can be made arbitrarily close to the actual detection cost function for a wide range of thresholds.
The minimum detection cost (minDCF) is achieved at a threshold where the DCF is minimized.
\begin{align}
    \text{minDCF} = \underset{\theta}{\min} \,\,C_{Norm}(\beta, \theta)
\end{align}
The threshold $\theta$ is included in the set of learnable parameters of the neural network. This way, the network learns to minimize the minDCF as a function of all the parameters through backpropagation.

\section{End-to-end modeling}
The model we explore is a concatenated version of two parameter tied x-vector extractors (TDNN networks~\cite{snyder2019speaker}) with the NPLDA model (Fig.~\ref{fig:e2e}). \footnote{The implementation of this model can be found in \url{https://github.com/iiscleap/E2E-NPLDA}}
The end-to-end model processes the mel frequency cepstral coefficients (MFCCs) of a pair of utterances to output a score. The MFCC features are passed through nine time delay neural network (TDNN) layers followed by a statistic pooing layer. The statistics pooling layer is followed by a fully connected layer with unit length normalization non-linearity. This is followed by a linear layer and a quadratic layer as a function of the enrollment and test embeddings to output a score. The parameters of the TDNN and the affine layers of the enrollment and test side of the E2E model are tied, which makes the model symmetric.

\subsection{GPU memory considerations}\label{ssec:gpu}
 We can estimate the memory required for a single iteration (batch update) of training as the sum of memory required to store the network parameters, gradients, forward and backward components of each batch. In this end-to-end network, each training batch of $N$ trials can have upto $2N$ unique utterances assuming there are no repetitions. For simplicity, let us assume each of the utterances corresponds to $T$ frames. We denote $k_i$ to be the dimension of the input to the $i^\text{th}$ TDNN layer, with a TDNN context of $c_i$ frames. The total memory required can then be estimated as
$2NT \sum_{i}{k_ic_i} \times 16 \text{ bytes.} $. The GPU memory is limited by the total number of frames that go into the TDNN, which is denoted by the factor $2NT$. A large batchsize of $2048$, as was used in \cite{ramoji2020leap}, is infeasible for the end-to-end model (results in GPU memory load of $240$GB). Hence, we resorted to a sampling strategy to reduce the GPU memory constraints. 
\subsection{Sampling of Trials}
\label{ssec:sampling}
In this current work, in order to avoid memory explosion in the x-vector extraction stage of the E2E model, we propose to use a small number of utterances ($64$) in a batch with about $20$ sec. of audio in each utterance. These $64$ utterances are drawn from $m$ speakers where $m$ ranges from $3-8$. These $64$ utterances are split randomly into two halves for each speaker to form enrollment and test side of trials. The MFCC features of the enrollment and test utterances are transformed to utterance embeddings $\eta _e$ and $\eta _t$ (as shown in Fig.~\ref{fig:e2e}). Each pair of enrollment, and test utterances is given a label as to whether the trial belongs to the target class (same speaker) or non-target class (different speakers). In this way, while the number of utterances is small, the number of trials used in the batch is $1024$.  Using the label information and the cost function  defined in Eq.~\ref{eq:det_cost}, the gradients are back-propagated to update the entire E2E model parameters.

This algorithm is applied separately to the male and female partitions of each training dataset to ensure the trials are gender and domain matched. All the $64$ utterances used in a batch come from the same gender and same dataset (to avoid cross gender, cross language trials). The algorithm is repeated multiple times with different number of speakers ($m$), for the male and female partitions of every dataset. Finally, all the training batches are pooled together and randomized. 

In contrast, the trial sampling algorithm used in our previous work on NPLDA \cite{ramoji2020nplda, ramoji2020leap} was much simpler. For each gender of each dataset, we sample an enrollment utterance from a randomly sampled speaker, and sample another utterance from either the same speaker or a different speaker to get a target or a non-target trial. This was done without any repetition of utterances, to ensure that each utterance appears once per sampling epoch. This procedure was repeated numerous times for multiple datasets and for both genders to obtain the required number of trials. All the trials were then pooled together, shuffled and split into batches of $1024$ or $2048$ trials.

It is worth noting that the batch statistics of the two sampling methods are significantly different. A batch of trials in the previous sampling method (Algo. 1) can contain trials from multiple datasets and gender, whereas in the modified sampling method, which we will refer as Algo. 2,  all the trials in a batch are from a particular gender of a particular dataset.


\section{Experiments and Results}
  The work is an extension of our work in \cite{ramoji2020leap}. The x-vector model is trained using the extended time-delay neural network (E-TDNN) architecture  described in \cite{snyder2019speaker}. This uses 10 layers of TDNNs followed by a statistics pooling layer. Once the network is trained, x-vectors of 512 dimensions are extracted from the affine component of layer 12 in the E-TDNN architecture. By combining the Voxceleb 1\&2 dataset \cite{nagrani2017voxceleb} with Switchboard, Mixer 6, SRE04-10, SRE16 evaluation set and SRE18 evaluation sets, we obtained with $2.2$M  recordings from $13539$ speakers. The datasets were augmented with the 5-fold augmentation strategy similar to the previous models. In order to reduce the weighting given to the VoxCeleb speakers (out-of-domain compared to conversational telephone speech (CTS)), we also subsampled the VoxCeleb augmented portion to include only $1.2$M utterances. The x-vector model is trained using $30$ dimensional MFCC features using a $30$-channel mel-scale filter bank spanning the frequency range $200$ Hz - $3500$ Hz,, mean-normalized over a sliding window of up to 3 seconds and with $13539$ dimensional targets using the Kaldi toolkit. More information about the model can be found in \cite{ramoji2020leap}. 
  
  The various backend PLDA models are trained on the SRE18 evaluation dataset. The evaluation datasets used include the SRE18 development and the SRE19 evaluation datasets. We perform several experiments under various conditions. The primary baseline to benchmark our systems is the Gaussian PLDA backend implementation in the Kaldi toolkit (GPLDA). The Kaldi implementation models the average embedding x-vector of each training speaker. The x-vectors are centered, dimensionality reduced using LDA to 170 dimensions, followed by unit length normalization. 

\pagebreak  
  In the traditional x-vector system, the statistic pooling layer computes the mean and standard deviation of the final TDNN layer. These two statistics then are concatenated into a fixed dimensional embedding. 
  We also perform experiments where we use variance instead of the standard deviation and compare the performance. 
  
  In the following sections, we study the influence of reduced training duration, and provide a performance comparison of the sampling method (Algo. 1 vs Algo. 2). We then compare the performance of Gaussian PLDA (GPLDA), Neural PLDA (NPLDA), and the proposed end-to-end approach (E2E). The PLDA backend training dataset used is the SRE18 Evaluation dataset. We report our results on the SRE18 Development set and the SRE19 Evaluation dataset using two cost metrics - equal error rate (EER) and minimum DCF ($C_{Min}$), which are the primary cost metrics for SRE19 evaluations.

\subsection{Influence of training utterance duration}\label{ssec:exp:dur}
As discussed in Section \ref{ssec:sampling}, due to GPU memory considerations and ease of implementation, we create a modified dataset by splitting longer utterances into 20 second chunks (2000 frames) after voice activity detection (VAD) and mean normalization. We compare the performances of the models on the modified dataset and the original one. The results are reported in Table \ref{tab:utt}. We observe that the performance of the systems are quite comparable. This allows us to proceed using these conditions in the implementation of the End-to-End model. All subsequent reported models use 20 second chunks for PLDA training.


\begin{table}[t]
    \centering
    \resizebox{\linewidth}{!}{
\begin{tabular}{@{}c|c|c|c|c|c@{}}
\toprule
\multirow{2}{*}{Model} & \multirow{2}{*}{\begin{tabular}[c]{@{}c@{}}Duration of \\ utterance \end{tabular}} & \multicolumn{2}{c|}{SRE18 Dev} & \multicolumn{2}{c}{SRE19 Eval} \\ \cmidrule(l){3-6} &  & EER (\%)      & $C_{Min}$ & EER (\%)& $C_{Min}$ \\ \midrule
GPLDA (G1) & Full & 6.43 & 0.417 & 6.18 & 0.512 \\
GPLDA (G2)   & 20 secs &  5.96  &0.436  & 5.80 & 0.518\\
\midrule
NPLDA (N1) & Full  &  5.33  & 0.389  & 5.10 & 0.443  \\
NPLDA (N2) & 20 secs & 5.57  & 0.359  & 5.32 & 0.432  \\ \bottomrule
\end{tabular}}
\vspace{0.25cm}
\caption{Performance comparison of training utterance durations (Full utterance vs 20 second segmenting) on GPLDA and NPLDA\cite{ramoji2020leap} models}
\vspace{-0.5cm}

\label{tab:utt}
\end{table}

\subsection{Comparison of sampling algorithms with NPLDA}
The way the training trials are generated is crucial to how the model trains and its performance. The performance comparison of the two sampling techniques with PLDA models trained on SRE18 Evaluation dataset can be seen in Table \ref{tab:sampling}. Although the nature of batch wise trials has changed significantly in terms of number of speakers in each batch and gender matched batches in the proposed new sampling method (Algo. 2), we see that its performance is comparable to our previous sampling method (Algo. 1).
\begin{table}[t]
    \centering
\resizebox{\linewidth}{!}{\begin{tabular}{@{}c|c|c|c|c|c@{}}
\toprule
\multirow{2}{*}{Model} &\multirow{2}{*}{Sampling} & \multicolumn{2}{c|}{SRE18 Dev} & \multicolumn{2}{c}{SRE19 Eval} \\ \cmidrule(l){3-6} 
&  & EER (\%)      & $C_{Min}$      & EER (\%)       & $C_{Min}$      \\ \midrule
NPLDA (N2) & Algo. 1                   & 5.57  & 0.359  & 5.32 & 0.432 \\
NPLDA (N3) & Algo. 2                   & 5.23  & 0.338  & 5.73   & 0.439 \\ 
\bottomrule
\end{tabular}}
\vspace{0.25cm}
\caption{Performance comparison with different sampling techniques using NPLDA\cite{ramoji2020leap} method using previous sampling method (Algo. 1) and proposed new sampling method (Algo. 2)}
\label{tab:sampling}
\end{table}

\subsection{End-to-End (E2E)}
 Using the proposed sampling method, we generate batches of 1024 trials using 64 utterances per batch. Both the NPLDA and E2E models were trained with this batch size. We use the Adam optimizer for the backpropagation learning. The performance of these models are reported in Table \ref{tab:e2e}. The NPLDA model is initialized with the GPLDA model. The initialization details of the models along with the pooling functions are reported in the table. We compare performances using two different statistics (StdDev or Var). We observe significant improvements in NPLDA over the GPLDA system and subsequently in E2E system over the NPLDA. Comparing E2E and GPLDA when we use standard deviation as the pooling function, we observe relative improvements of about $23$\% and $22$\% in SRE18 development and SRE19 evaluation sets, respectively in terms of the $C_{Min}$ metric. The relative improvements between E2E and GPLDA when we use Var as the pooling function are about $33$\% and $20$\% for SRE18 development and SRE19 evaluation sets, respectively for the $C_{Min}$ metric. Though, the cost function in the neural network aims to minimize the detection cost function (DCF), we also see improvements in the EER metric using the proposed approach. These results show that the joint E2E training with a single neural pipeline and optimization results in improved speaker recognition performance. 

\begin{table}[t!]
    \centering
    \resizebox{\linewidth}{!}{
\begin{tabular}{@{}l|c|c|c|c|c|c@{}}
\toprule
\multirow{2}{*}{Model} & \multirow{2}{*}{\begin{tabular}[c]{@{}c@{}}Pooling \\ function \end{tabular}} & \multirow{2}{*}{Init.} & \multicolumn{2}{c|}{SRE18 Dev} & \multicolumn{2}{c}{SRE19 Eval} \\  \cmidrule(l){4-7} 
 & \multicolumn{1}{l|}{} &  & EER (\%) & $C_{Min}$ & EER (\%) & $C_{Min}$ \\ \midrule
GPLDA (G2) & StdDev & - & 5.96 & 0.436 & 5.80 & 0.518 \\
GPLDA (G3) & Var & - & 7.23 & 0.459 & 6.33 & 0.560 \\
NPLDA (N2) & StdDev & G2 & 5.57 & 0.359 & 5.32 & 0.432 \\
NPLDA (N4) & Var & G3 & 6.05 & 0.377 & 5.91 & 0.465 \\
E2E (E1) & StdDev & N2 & \textbf{5.36} & 0.337 & \textbf{5.31} & \textbf{0.405} \\
E2E (E2) & Var & N4 & 5.60 & \textbf{0.307} & 5.43 & 0.446 \\ \bottomrule
\end{tabular}}
\vspace{0.25cm}
\caption{Performance comparison between GPLDA, NPLDA and E2E models using standard deviation and variance as the secondary pooling functions. The model that was used to initialize the network is denoted in the 3rd column}
\label{tab:e2e}
\vspace{-0.5cm}
\end{table}

\section{Summary and Conclusions}
This paper explores a step in the direction of a neural End-to-End (E2E) approach in speaker verification tasks. It is an extension of our work on a discriminative neural PLDA (NPLDA) backend. The proposed model is a single elegant end-to-end approach that optimizes directly from acoustic features like MFCCs with a verification cost function to output a likelihood ratio score. We discuss the influence of the factors that were key in implementing the E2E model. This involved modifying the duration of the training utterance and developing a new sampling technique to generate training trials. The model shows considerable improvements over the generative Gaussian PLDA and the NPLDA models on the NIST SRE 2018 and 2019 datasets.  One drawback of the proposed method is the requirement to initialize the E2E model with pre-trained weights of an x-vector network. 

Future work in this direction could include investigating better sampling algorithms such as the use of curriculum learning \cite{ranjan2017curriculum}, different loss functions, improved architecture for the embedding extractor using attention and other sequence models such as LSTMs etc.

\bibliographystyle{IEEEtran}

\bibliography{template}


\end{document}